\documentclass[conference, 10 pt]{ieeeconf}

\IEEEoverridecommandlockouts                    
\usepackage{amsmath,amsfonts}
\usepackage{algorithmic}
\usepackage{algorithm}
\usepackage{array}
\usepackage{textcomp}
\usepackage{stfloats}
\usepackage{url}
\usepackage{verbatim}
\usepackage{graphicx}
\usepackage{cite}
\usepackage{multirow}

\begin{document}
\title{Deep Neural Network NMPC for Computationally Tractable Optimal Power Management of Hybrid Electric Vehicles}
\author{Suyong Park$^{1}$, Duc Giap Nguyen$^{1}$, Jinrak Park$^{2}$, Dohee Kim$^{2}$, Jeong Soo Eo$^{2}$, and Kyoungseok Han$^{1*}$
\thanks{$\star$ This work was supported in part by Hyundai Motor Company; in part by NRF funded by the Korean Government [Ministry of Science and ICT (MSIT)] (NRF-2021R1C1C1003464); in part of Basic Science Research Program through the National Research Foundation of Korea (NRF) funded by the Ministry of Education (2021R1A6A1A03043144).}
\thanks{$^{1}$S. Park, D. Nguyen, and K. Han are with Mechanical Engineering,
        University of Kyungpook National, 41566 Buk-gu, Daegu, Republic of Korea 
        {\tt\small \{suyongpark, everless95, kyoungsh\}@knu.ac.kr}}%
\thanks{$^{2}$J. Park, D. Kim, and J. Eo are with Hyundai Motor Company, 18278 Namyang-eup, Hwaseong-si, Republic of Korea
        {\tt\small \{pjr, doheekim, fineejs\}@hyundai.com}}%
}

\maketitle

\begin{abstract}
This study presents a method for deep neural network nonlinear model predictive control (DNN-MPC) to reduce computational complexity, and we show its practical utility through its application in optimizing the energy management of hybrid electric vehicles (HEVs). For optimal power management of HEVs, we first design the online NMPC to collect the data set, and the deep neural network is trained to approximate the NMPC solutions. We assess the effectiveness of our approach by conducting comparative simulations with rule and online NMPC-based power management strategies for HEV, evaluating both fuel consumption and computational complexity. Lastly, we verify the real-time feasibility of our approach through process-in-the-loop (PIL) testing. The test results demonstrate that the proposed method closely approximates the NMPC performance while substantially reducing the computational burden.
\end{abstract}


\section{Introduction}
With communication technology advancements, such as V2X, researchers have dedicated efforts to devising energy-efficient driving strategies that conserve energy and mitigate emissions.
One promising avenue for achieving this efficiency is the exploitation of predicted traffic information. The goal of this work is to maximize fuel economy, we focus herein on hybrid electric vehicles (HEVs) and the development of energy-efficient driving targeting the optimal power distribution of energy resources \cite{b3}. An appropriate power distribution strategy will enable us to achieve the optimal energy distribution between an internal combustion engine and an electric motor.

Dynamic programming (DP) is generally employed to solve optimal control problems \cite{b4}. However, while it can ensure both global optimality and constraint enforcement, its computational load is a challenge when achieving real-time implementation. The indirect approach of Pontryagin's minimum principle (PMP) is a widely adopted alternative focused on minimizing fuel usage \cite{b5, b6}. This approach transforms the optimal control problem into a boundary value problem. While the PMP method can obtain optimal solutions that closely approximate the outcomes produced by DP, its computational complexity in solving optimal control problems makes it impractical for direct analytical resolution. Consequently, effective solutions to the ensuing boundary value problem necessitate the application of approximation techniques. 

Designed to solve optimal control problems, model predictive control (MPC) is one of the promising approaches proposed to handle the drawbacks of DP and PMP \cite{b7, b8}. MPC has gained widespread adoption in both academia and industry due to its proven superior performance. However, dependence on techniques, such as linearization and reformulation of the quadratic cost function, which is necessary for real-time implementation, may compromise its effectiveness. Numerous studies have demonstrated the effectiveness of MPC in reducing the energy consumption of HEVs \cite{b9, b10}. These solutions guarantee optimal results, but the critical question as regards their computational time requirements for real-time applications remains. Achieving a strong performance is essential, but equally crucial is the assessment of their feasibility for real-world implementation. The powertrain model of an HEV is especially highly nonlinear; hence, the computational complexity of MPC significantly increases as compared with the linear system application. Although nonlinear MPC has a superior performance, the computational burden complicates its real-time application.

Motivated by the abovementioned drawback, we propose herein a nonlinear MPC (NMPC) approximation solution based on supervised deep learning (DNN-MPC) for real-time optimization in HEVs targeting repetitive daily routines. DNN-MPC maintains performance while significantly reducing computational burdens by evoking the optimal control sequence \cite{b11}. We design an NMPC approximation, which yields a superior performance over the conventional rule-based approach, to facilitate the DNN-MPC learning. Their real-world applicability is evaluated through various simulations and process-in-the-loop (PIL) experiments conducted with a high-fidelity powertrain model.

\section{\uppercase{System Modeling and PROBLEM FORMULATION}}
\subsection{Control-oriented Model}

\begin{figure}[t!]
\centering
    \includegraphics[width=75mm]{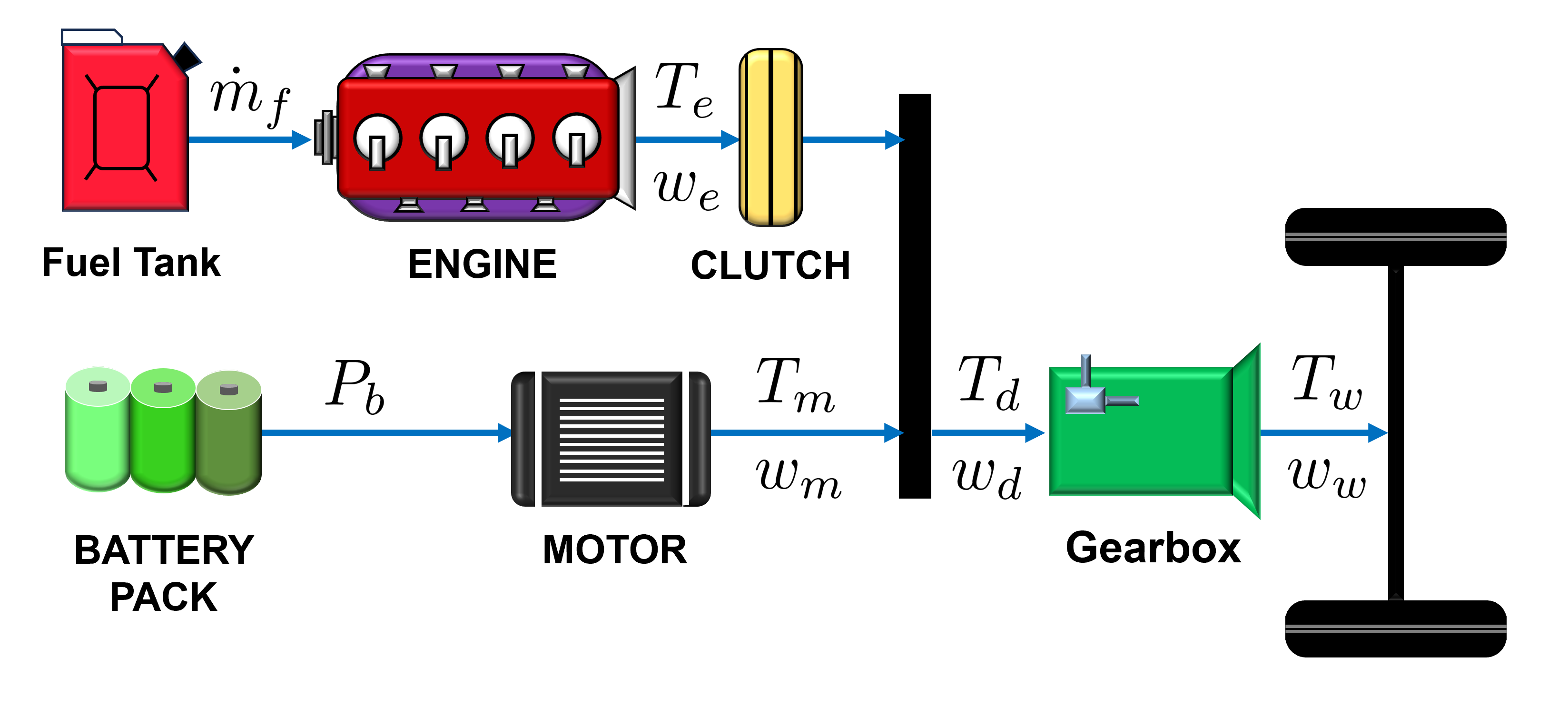}
    \caption{Power flows in the parallel and mild HEVs:  $\dot{m}_f$ is fuel consumption; $P_b$ the battery power; $T$ the torque; $w$ the rotational speed; and subscripts $e, m, d, \text{and }, w$ are engine, motor, demand, and wheel respectively.}
    \label{HEV_conf}
\end{figure}

In this work, we considered parallel and mild HEVs \cite{b12}, that blend a traditional fuel engine and electric power with a small battery capacity (Fig.~\ref{HEV_conf}).

The demanded torque $T_d$ determined by the vehicle speed profile $v$ is computed as follows:
\begin{align}
    T_d(v) = B\cdot \underbrace{T_e(u)}_{=uT_d}+\underbrace{T_m(u)}_{=(1-u)T_d}
    \label{torque_eq}
\end{align}
where $T_e$ and $T_m$ are the engine and motor torques respectively; $u\in[0,1]$ is the torque distribution ratio; and $B$ = $\left \{    1,0  \right \}$ is the engine status. When the engine is engaged, $B=1$; otherwise, $B=0$. The emphasis was not on the mode change. hence, we set $B=1$ when the distribution ratio $u=0$.

The engine fuel rate $\dot{m}_f(T_e, w_e)$ and the battery power $P_b(T_m, w_m)$ are computed as follows using pre-established polynomial functions \cite{b13}:
\begin{subequations}
\begin{align}
&\dot{m}_{f} =  a_{0}+a_{1}T_{e}+a_{2}w_{e}+a_{3}T_{e}^2
    +a_{4}T_{e}w_{e}+a_{5}w_{e}^2,   \\ 
&P_{b} = b_{0}+b_{1}T_{m}+b_{2}w_{m}+b_{3}T_{m}^2
    +b_{4}T_{m}w_{m}+b_{5}w_{m}^2.
\end{align}
\end{subequations}
where $a_i$ and $b_i$, $i\in [0:5]$ are coefficients, and $w_e$ and $w_m$ are calculated from the vehicle speed profiles. A more detailed HEV model is presented in \cite{b14}.

To represent the evolving charging and discharging dynamics of the battery, the state $x$ is the battery state-of-charge ($SOC$). Using Euler discretization, state ($x=SOC$) can be expressed by:
\begin{equation} 
\small
    x(t+1) = x(t)+\frac{V_{c}\left( x,t\right) - \sqrt{V_{c}^2 \left( x,t\right) - 4R_{b}\left( x,t\right)P_{b}\left( x,t\right)}}{2R_{b}\left( x,t\right)C}T_s
    \label{SOC_dyn}
\end{equation}
\normalsize
where $x=SOC$ is state; $T_s$ is sampling time; $V_{c}$ is the open-circuit voltage; $R_{b}$ is the battery resistance; and $C$ is the battery maximum capacity.

\subsection{Nonlinear MPC Design}
\begin{figure}[t!]
\centering
    \includegraphics[width=80mm]{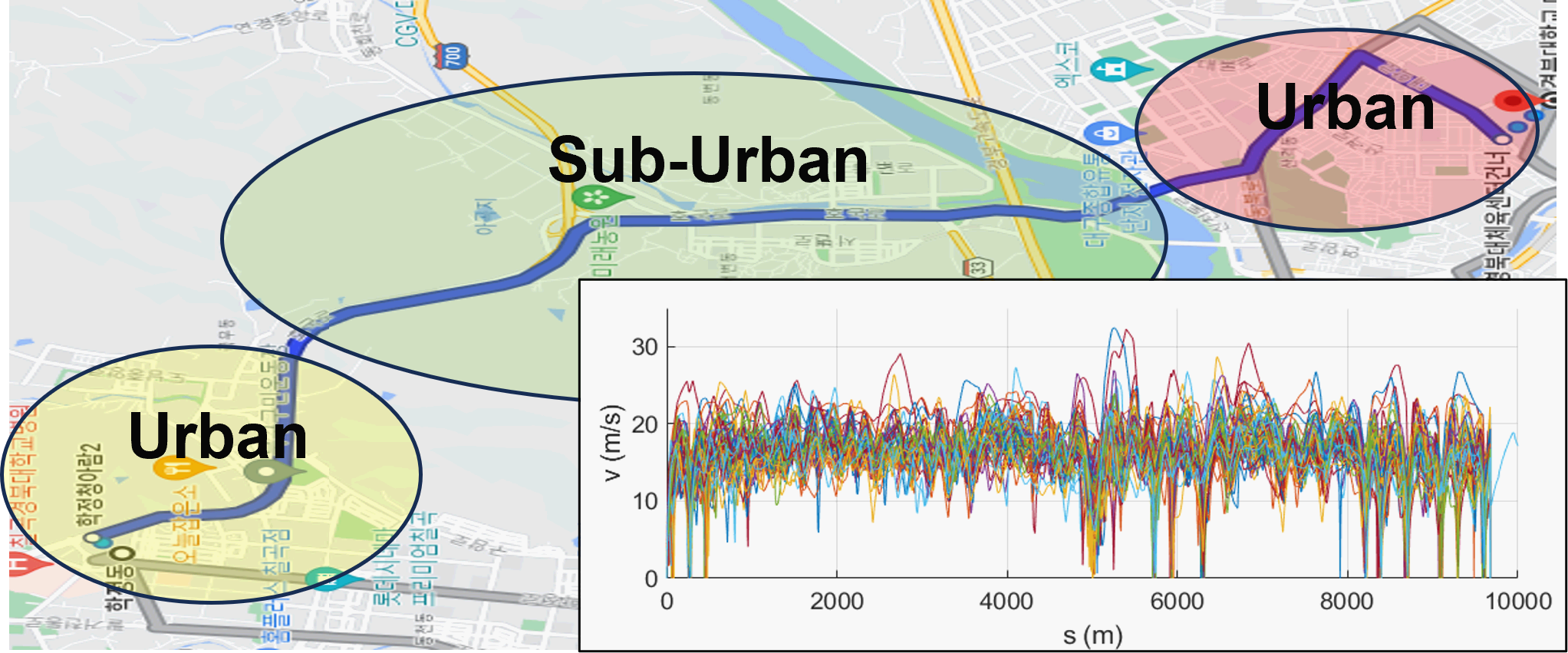}
    \caption{Multiple speed profiles obtained by the driving simulator from Hakjeong-dong to Kyungpook National University in Daegu (9.6 km)}
    \label{speed profile}
\end{figure}

As shown in Fig.~\ref{speed profile}, we repeatedly collected the multiple speed profiles through driving simulations for a specific route, making the speed profiles similar to each other \cite{b15}.

We assumed that the predicted acceleration sequence for the next $N$ steps $[a(t), a(t+1), \cdots, a(N-1)]$ is available through vehicle connectivity technologies; thus, the demanded torque sequence $T_d$ over the prediction horizon $N$ can also be predicted. 

The optimal control problem for the optimal torque distribution is now formulated as
\begin{subequations} \label{NMPC}
\begin{align}
   & \min_{u(k),\cdots,u(k+N-1)} \sum^{N-1}_{k= 0}  \Delta m_{f}(u(k | t)) + \lambda(k) \cdot \Delta SOC(u(k | t)) \nonumber  \\
    & \quad \text{subject to} \nonumber \\
    & x(k+1 | t) = x(k | t) + f(x(k |t),u(k | t))T_{s} \\
    & x^{\mathrm{min}} \leq x(k+1 | t) \leq x^{\mathrm{max}}  \\
    & 0 \leq u(k|t) \leq 1 \\
    & k = 0,1, \cdots, N-1 \nonumber
\end{align}
\end{subequations}
where $\Delta{m_f}=\dot{m}_f T_s$ denotes the fuel consumption; control input $u$ is torque distribution ratio; $\lambda(k)$ refers to the equivalent factor known as the dimensionless conversion ratio, quantifying the conversion of electrical power into chemical power flow; $N$ is the prediction horizon; and $T_s$ is the sampling time.

For the energy-efficient management mild HEVs, the goal is to minimize the fuel consumption while maintaining the battery $SOC$ within a predefined range. The equivalent factor $\lambda(k)$ can be adjusted to maintain the battery $SOC$ within the predefined constraints. Similar driving patterns were observed for the same route (Fig.~\ref{speed profile}) at a specific period; therefore, we obtain the $\lambda(k)$ value based on the average vehicle speed of multiple cycles using DP.

\begin{figure}[t!]
\centering
    \includegraphics[width=85mm]{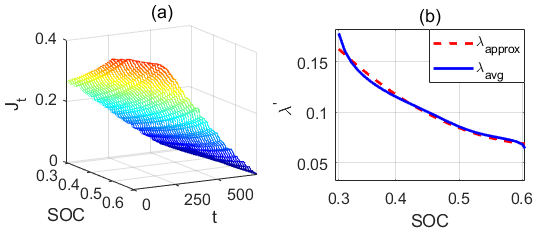}
    \caption{Cost-to-go for the average driving outcomes: (a) DP cost-to-go value and (b) approximated $\lambda_k$ from Eq.~\eqref{app_lambda}.}
    \label{DPCOST}
\end{figure}

First, the Bellman's principle of optimality is used \cite{b16}: 
\begin{subequations}
\begin{align}
    T^*_e (u^*) &= \underset{u}{\text{argmin}}\left [ \dot{m}_{f,t}(u)T_s +\mathcal{J}_{t+1}(SOC_{t+1}) \right ]  \\ \nonumber
    &\approx \underset{u}{\text{argmin}} \left [
\dot{m}_{f,t}(u)T_s + \mathcal{J}_t(SOC_t)  \cdots \right.\\ 
    & \left. \quad + \frac{\partial \mathcal{J}_t(SOC_t)}{\partial t}T_s + \frac{\partial \mathcal{J}_t(SOC_t)}{\partial SOC} f(x,u)T_s \right ] \\ 
   & =  \underset{u}{\text{argmin}} \left [ \Delta{m}_{f,t}(u) + \lambda_t^{'} \Delta f(x,u) \right ]
\end{align}
\end{subequations}
where $\mathcal{J}(SOC_t)$ is the cost-to-go function of the battery $SOC$ at time $t$, and  $\lambda_t^{'}$ = $\partial \mathcal{J}_t(SOC_t)$/$\partial SOC$.  
To calculate $\lambda_t^{'}$, the functional $\mathcal{J}_t(SOC_t)$ that is dependent on the control $u_t^*(SOC_t)$ and $SOC_t$ can be obtained by DP \cite{b17}, as shown in Fig.~\ref{DPCOST} (a). Since $\lambda_t^{'}$ is time-varying, we substitute $\lambda_t^{'}$ with time-averaged value $\lambda_{avg}^{'}$ as following:
\begin{align}
    \lambda_{avg}^{'} = \frac{1}{N_c} \sum^{N_c}_{t=0} \lambda_t^{'}, 
    \label{app_lambda}
\end{align}
where $N_c$ is the cycle time. Fig.~\ref{DPCOST}(b) presents the averaged $\lambda_{avg}^{'}$. Utilizing Fig.~\ref{DPCOST}(b) as a lookup table will increase the computation time; hence, this was accomplished by substituting the conventional constant equivalent factor with an expression that considers into account both the current state-of-charge $SOC(t)$ and the desired reference value $SOC_{\text{ref}}$.
\begin{align}
    \lambda_{approx}^{'} \approx \lambda_0 + K_a(SOC(k)-SOC_{\text{ref}})^2,
    \label{lambda_avg}
\end{align}
where $\lambda_0$ and $K_a$ are the tuning parameters. The reference value $SOC_{\mathrm{ref}}$ is equated to the initial $SOC_{init}$ to realize the CS mode. The Eq.~\eqref{lambda_avg} can be utilized in the equivalent factor $\lambda(k)$ in 
 the cost of Eq.~\eqref{NMPC}.

\subsection{NMPC Control Result}
\begin{figure}[t!]
\centering
    \includegraphics[width=80mm]{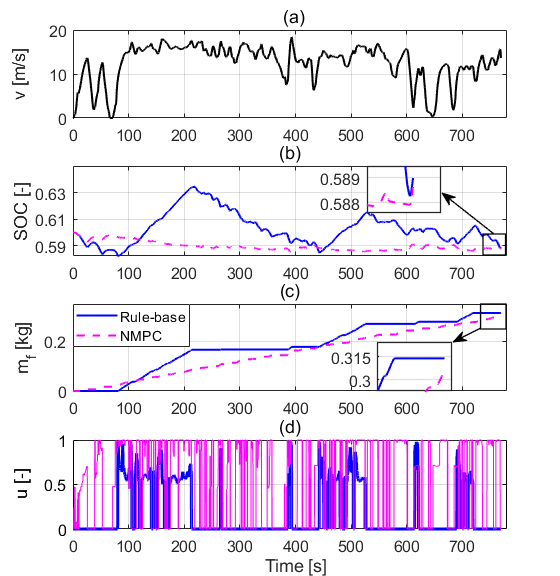}
    \caption{Simulation results: (a) speed profile, (b) SOC trajectory, (c) cumulated fuel consumption, and (d) power distribution ratio. }
    \label{implicit_design}
\end{figure}

Fig.~\ref{implicit_design} compares the control performances of rule-based and NMPC approaches. The rule-based control comes from the Autonomie commercial software \cite{b18}, and its fundamental principle is a test-based map created from test results. The speed profile was generated through driving simulations for a specific route (Fig.~\ref{implicit_design}(a)). Fig.~\ref{implicit_design}(b) shows both strategies approaching the reference value $(SOC_{ref}=0.6)$ with the CS mode. NMPC achieves lower fuel consumption compared to the rule-based strategy, despite both approaches achieving identical SOC levels (see Fig.~\ref{implicit_design}(c)). It yielded a 4.76\% improvement in terms of fuel consumption, as compared to the rule-based approach and this is possible with the predicted traffic information.

\begin{figure}[t]
\centering
    \includegraphics[width=80mm]{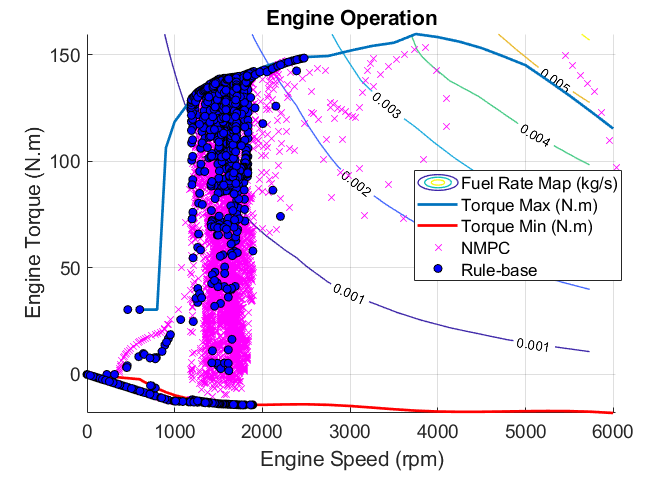}
    \caption{Engine operation points of the rule-based and NMPC methods.}
    \label{evaluation_implicit_operation}
\end{figure}

Fig.~\ref{evaluation_implicit_operation} compares the engine operation points $\dot{m}_{f}$ based on the engine torque and speed. A notable distinction emerged between the majority of the operation points for the rule-based strategy, which tended to cluster around $0.002\,\text{kg}/\text{s}$, and those for the NMPC strategy, which were distributed closer to $0.001\,\text{kg}/\text{s}$. Importantly, the rule-based approach exhibited the sudden spike tendency in engine power consumption. Despite achieving the same final SOC levels, this power distribution inefficiency led to a higher fuel consumption compared to NMPC.

\section{Approximated NMPC using Neural Network}

In this part, we present the utilization of DNN for approximating NMPC. The process of NMPC approximation involves three main stages: data collection, DNN training, and evaluation of optimality.
\subsection{DNN Training Procedure}
The supervised learning approach aimed to determine the weight factors denoted as $\textbf{W}_{i}$.

Given the input data set $\textbf{X}$ and the corresponding output data set $\mathbf{\hat{Y}}$, the NN underwent training to approximate the NMPC results. We employed a feed-forward NN to obtain the weight factors. The training steps for this are as follows:

\textbf{1) Speed profiles collection:} We collected various speed profiles from multiple driving cycles for a specific route (Fig.~\ref{speed profile}). The demanded acceleration of the vehicle sequence was also obtained: $\{a(0), \cdots, a(t_f-1)\}$ where $t_f$ is the final time of a specific driving cycle.

\textbf{2) NMPC data collection:} In the NMPC formulation in Eq.~\eqref{NMPC}, the optimal control sequence was solved given the short-term preview vehicle acceleration demand sequence and the battery $SOC$ at the current time step. We approximated this NMPC collecting various pairs of state and control from the NMPC. The input and output data sets to NN are defined as follows: 
    \begin{subequations}
        \begin{align}
            &\textbf{X}= \left \{SOC(k), [\underbrace{ a(k), a(k+1), \cdots,  a(k+N-1)]}_{\text{Short-term predicted future acceleration sequence}} \right \}, \\
            &\hat{\textbf{Y}} = \left \{ \hat{u}(k), \hat{u}(k+1), \cdots, \hat{u}(k+N-1) \right \}.
        \end{align}
    \end{subequations}
    \normalsize
    
    where $\textbf{X}$ and $\hat{\textbf{Y}}$ are the input and output data set to NN respectively.
\begin{table}[t!]
\vspace{5mm}
\caption{DNN Hyperparameters}
\centering
\small
\renewcommand{\arraystretch}{1.5}
{%
\begin{tabular}{lr}
\hline
\textbf{Hyperparameters}             & \textbf{Values} \\ \hline\hline
Training times for one NN & 3     \\ 
Maximum structure change   & 4     \\ 
Number of Data set         & 7,500 \\ 
Number of layers of NN     & 3     \\ 
Epochs                     & 500   \\ 
Initial neurons within each layer  & $\left \{  5,5,3 \right \}$ \\ \hline
\end{tabular}%
}
\label{hyperparameters}
\end{table}
\begin{figure}[b]
\centering
    \includegraphics[width=85mm]{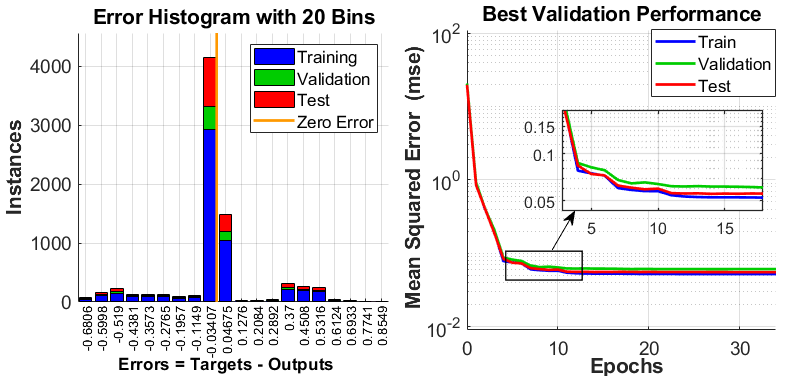}
    \caption{DNN-MPC training performance: (a) error histogram and (b) best validation performance of the final neural network. }
    \label{Train_results}
\end{figure}
\textbf{3) Training parameters:} The optimal parameter values of $\mathbf{W}_i$ were identified through the Marquardt algorithm based on backpropagation using the collected data set \cite{b19}. The process of obtaining the approximated output $\textbf{Y}^*$ is expressed as follows:
    \begin{align}
             \textbf{Y}^* = p_{l} \circ \sigma  \circ p_{l-1} \circ \sigma \circ \cdots \circ  p_{1} \circ \sigma
    \end{align}
    where $p = \textbf{W}_{i}\textbf{X} + \textbf{b}_{i}$ is affine function, $\textbf{b}_{i}$ is bias, $l$ is the number of neuron, and $\sigma$ is activation function. 

\textbf{4) DNN structure determination:} Given a fixed number of layers, the number of neurons inside each layer was determined by trial-and-error. Starting from the initially given numbers of $\left \{5, 5, 3\right \}$, the numbers were incrementally expanded until the optimal training performance was achieved. The training performance is assessed as follows:
    \begin{align}
        e = \textbf{Y}^*(\textbf{X}) - \hat{\textbf{Y}}(\textbf{X})
        \label{error}
    \end{align}
    where $e$ is the error between $\textbf{Y}^*(\textbf{X})$ and $\hat{\textbf{Y}}(\textbf{X})$ generated by neural network and NMPC respectively. 

    Table~\ref{hyperparameters} shows the determined DNN hyperparameters for the specific problem presented in this work. The NN structure had three layers, and the initial NN $\left \{5, 5, 3\right \}$ was expanded by four times. One NN was trained by thrice. One per training time involved repeating 500 epochs to obtain the training parameters. The hyperparameters were obtained by trial and error until the best training performance was achieved.

\subsection{Training Validation}
The NN for the energy distribution was trained using the MATLAB function. Fig.~\ref{Train_results} summarizes the learning results based on 7500 data sets. The available data were divided into three subsets. The first subset was used as the training set to compute the gradients and update the neural network weights and biases. The second subset served as the validation set, and its error was monitored during the training process. The third subset was the test set, and its error is monitored after the training completion.

Fig.~\ref{Train_results}(a) illustrates the error histogram calculated using Eq.(\ref{error}) for the high-performance neural network. Most data sets were situated near zero error with the minimum error between $\textbf{Y}^*(\textbf{X})$ and $\hat{\textbf{Y}}(\textbf{X})$. In Fig~\ref{Train_results}(b), the training error converged to a minimal level confirming the hyperparameter selection (Table~\ref{hyperparameters}).

\section{Test Results}

\subsection{Test Environments}
To validate our proposed DNN-MPC, we assessed its performance by comparing it with NMPC and a rule-based approach. Fig.~\ref{final_results}(a) shows the speed profiles divided into trained and untrained data sets for DNN-MPC. Trained data represents the data utilized for the training of DNN-MPC, while untrained data was employed to evaluate the performance, as it was not used during the training process. The shaded area corresponds to $\pm$1.5 times the region of the trained data, and the data outside the region exhibits decreased similarity with the trained data, which shows that the speed profiles can be different even when the vehicle is travelling the same route. We evaluated the performance and computation time of DNN-MPC based on both trained and untrained data sets.

\subsection{Simulation Results}

\begin{figure}[!t]
\vspace{5mm}
\centering
    \includegraphics[width=80mm]{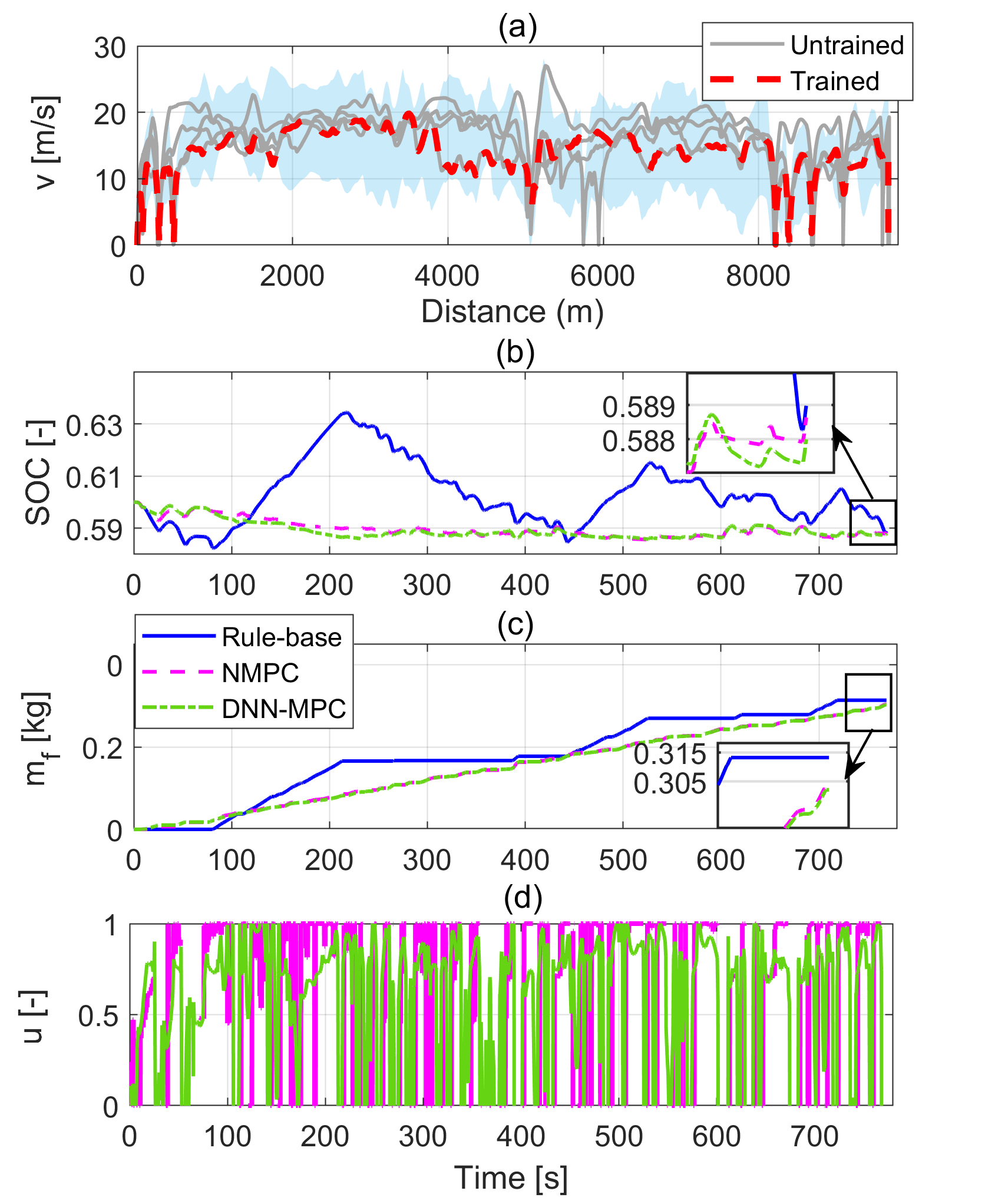}
    \caption{Comparison of the simulation results of different three approaches: (a) speed profile; (b) SOC trajectory; (c) fuel consumption; (d) control input.}
   \label{final_results}
\end{figure}

 The similarity between NMPC and DNN-MPC was assessed based on the trained data, as depicted in Fig.~\ref{final_results}. both
strategies approach the reference value ($SOC_{ref}$ = 0.6)
with the CS mode. However, in Fig.~\ref{final_results}(c), NMPC and DNN-MPC showed lower fuel consumption compared to the rule-based approach. DNN-MPC approximated the NMPC control policy (Fig.~\ref{final_results}(d)). thereby exhibiting a fuel consumption similar to that of NMPC.

\begin{figure}[t!]
\centering
    \includegraphics[width=80mm]{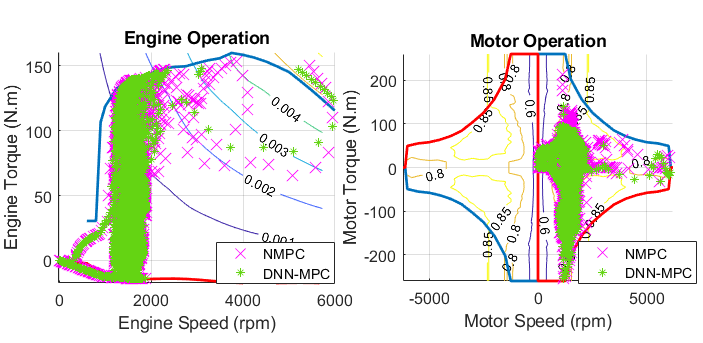}
    \caption{Comparison of engine and motor operating points of NMPC and DNN-MPC for mitigating the NN overfitting.}
    \label{oeparation_eninge_mot}
\end{figure}

\begin{figure}[t!]
\vspace{5mm}
\centering
    \includegraphics[width=80mm]{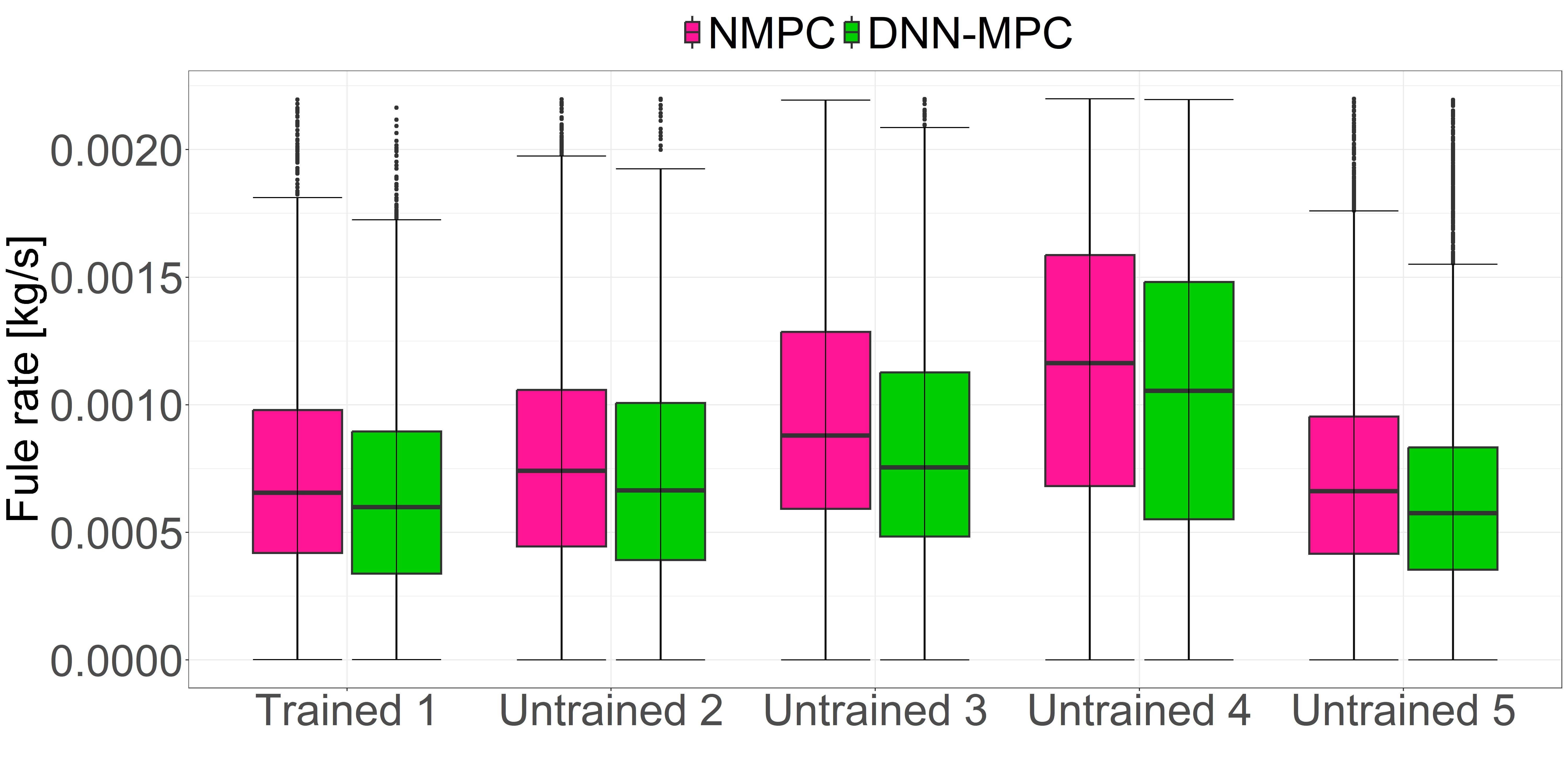}
    \caption{Fuel flow rate similarities among three methods.}
    \label{boxplot}
\end{figure}

 Fig.~\ref{oeparation_eninge_mot} depicts the operation points of the engine and the motor. Based on the trained data, the operation points of NMPC and DNN-MPC almost overlapped. To assess how closely DNN-MPC approximated NMPC, we conducted a statistical analysis using the results of Fig.~\ref{boxplot}. Fig.~\ref{boxplot} illustrates the distribution of the fuel rate operating points with the data sets divided into the trained and untrained categories. 
The median values of NMPC and DNN-MPC in Trained 1 recorded a 3.4\% deviation, indicating their close resemblance. The untrained speed profile showed minimum and maximum differences of 10.2\% and 14.2\% respectively. Despite using untrained data sets in the simulation, DNN-MPC generated similar control inputs to NMPC on specific repeated routes.

\begin{table}[t!]
    \caption{SIMULATION RESULTS}
    \label{total_cost_table}
\begin{minipage}[t]{.3\textwidth}
\centering
\setlength{\tabcolsep}{5pt}
\renewcommand{\arraystretch}{1.3}
\small{%
\begin{tabular}{cccccccc}
\hline
\multicolumn{1}{c}{Data} &
\multicolumn{1}{c}{Strategy} &
\multicolumn{1}{c}{$SOC_f(-)$} &
\multicolumn{1}{c}{$m_{f}$(kg)} &
\multicolumn{1}{c}{Cost(\$)}  \\ \hline
\multirow{3}{*}{Trained 1} & Rule-based  & 0.589   & 0.313  & 1.733         \\
                       &NMPC & 0.588   & 0.303 & 
                       1.695  \\ 
                        & DNN-MPC & 0.588    & 0.302 & 1.690  \\ 
                         \hline
\multirow{3}{*}{Untrained 2} & Rule-based  & 0.577 & 0.311  & 1.738      \\
                       &NMPC & 
                       0.597 &
                       0.333 &
                      1.798   \\ 
                        & DNN-MPC & 0.588    & 0.319 &1.760  \\
                         \hline
\multirow{3}{*}{Untrained 3} & Rule-based  & 0.573   & 0.403   & 2.111        \\
                       & NMPC & 
                       0.572  & 
                       0.394& 
                      2.074    \\ 
                        & DNN-MPC &0.584    &0.401 & 2.085  \\
                         \hline
\end{tabular}}
\end{minipage}
\end{table}

Table~\ref{total_cost_table} presents the results of the battery SOC, fuel consumption, and associated energy consumption cost. The fuel and battery costs were assumed as $3.954~US\$/kg$, and $0.169~US\$/kwh$ obtained by the United States Energy
Information Administration respectively \cite{b20}.
For the trained data sets, the NMPC and DNN-MPC recorded fuel consumption of 0.303 kg and 0.302 kg respectively. The rule-based strategy recorded 0.313 kg, resulting in a 2.19\% higher cost compared to the NMPC. However, The untrained data sets revealed the overall differences between NMPC and DNN-MPC. Untrained 1 depicted 1.50\% difference in the SOC and a 4.20\% difference in fuel consumption between NMPC and DNN-MPC. As aforementioned, the driving speeds for the repeated routes were similar (Fig.~\ref{speed profile}). In other words, DNN-MPC can provide coverage using the trained data set. Untrained 2 demonstrated a 1.88\% difference in the SOC and a fuel consumption difference of 0.49\% between DNN-MPC and NMPC. Compared with the rule-based strategy, it consumed the least fuel in Untrained 1 because it required higher battery usage. By contrast in Untrained 2, DNN-MPC demonstrated superiority by achieving the almost same fuel consumption as the rule-based strategy while saving approximately 1.88\% of the battery.

\subsection{Process-in-the-Loop Experiments}
\begin{figure}[htb!]
\centering
    \includegraphics[width=85mm]{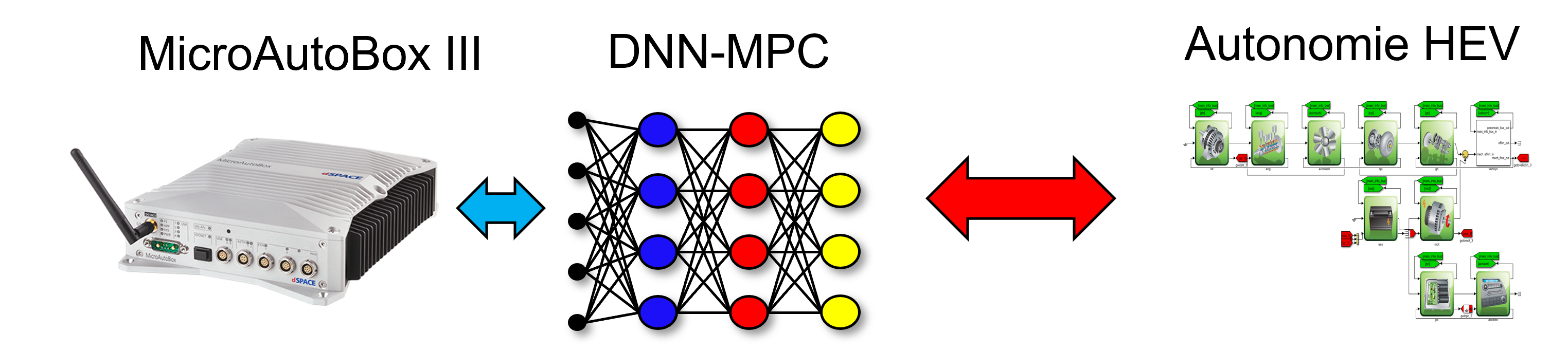}
    \caption{Process-in-the-loop experimental architecture.}
    \label{PIL}
\end{figure}

We validated the computation time through the PIL test. (Table~\ref{comp_time}). As shown in Fig.~\ref{PIL}, the developed controllers were embedded in MicroAutoBox III, which is a dual-core ARM Cortex-A9 processor. The clock frequency was ensured up to 2 x 1.5 GHz with 32 GB eMMC onboard flash memory. The rule-based approach demonstrated a minimum calculation speed of 0.053ms, a maximum of 0.062ms, and an average calculation speed of 0.055ms. Meanwhile, NMPC exhibited an average calculation speed of 18.32ms, which did not meet the production-standard control speed of 10ms. NMPC may even reach a maximum computing speed of 178.67ms at times, making it unsuitable for real-time applications. By comparison, DNN-MPC achieved a near-optimal performance similar to NMPC but with the advantage of a real-time application. DNN-MPC exhibited an average computation speed of 0.068ms. Although DNN-MPC demonstrated a slightly slower average computational speed compared to the rule-based approach, it remained well-suited for real-time application. On the contrary, NMPC showed a considerably slower computational capability that makes real-time applications infeasible. The PIL experiment verified the real-time capability and capacity of DNN-MPC to approximate optimality at a level comparable to that of NMPC.

\begin{table}[t!]
\caption{Computation time evaluation }
\centering
\renewcommand{\arraystretch}{1.3}
\small
\begin{tabular}{c|ccc}
\hline
\multirow{2}{*}{Methods} & \multicolumn{3}{c}{Computation time (ms)} \\ \cline{2-4} 
                         & Min        & Avg        & Max        \\ \hline
Rule-based approach                 & 0.053         & 0.055          & 0.062          \\ \hline
NMPC                      & 14.80          & 18.32          & 178.67          \\ \hline
DNN-MPC                   & 0.063          & 0.068          & 0.072          \\ \hline
\end{tabular}
\label{comp_time}
\end{table}

\subsection{Limitations}
While we initially assumed that HEVs followed specific recurring routes, we encountered instability in their control performance when untrained data sets, diverging from the expected speed profile range, were introduced into the DNN-MPC system. In this system, DNN-MPC approximated the control inputs for NMPC based on the training data sets ${a(0), \cdots, a(t_f-1)}$. As a result, untrained data led to the accumulation of errors arising from uncertainty, thereby impairing the approximation of the NMPC control policy. This deficiency in robustness was particularly pronounced when dealing with untrained speed profiles. To address this challenge, our forthcoming research endeavors will encompass the incorporation of a stochastic traffic model, such as the Markov traffic model, and its integration into our control design methodology.

\section{Conclusion}
We presented herein an energy-efficient control strategy for mild HEVs that aims to minimize fuel consumption. We designed NMPC for the data set collection and proposed a deep neural network-based approximation to reduce the computational time. DNN-MPC demonstrated superiority by achieving the almost same fuel consumption as the rule-based strategy while saving approximately 1.88\% of the battery.
We also assessed the real-time capability of DNN-MPC by conducting PIL experiments. The test
results demonstrate that the proposed method closely approximates the NMPC performance while substantially reducing the computational burden.

\end{document}